\documentclass[12pt,letter]{article}
\usepackage{amssymb}
\usepackage{amsmath}
\usepackage{graphicx}
\usepackage{cite}
\usepackage{float}
\usepackage{caption}
\usepackage{authblk}
\usepackage{subfigure}
\usepackage[margin=1.0in]{geometry}
\usepackage{color}

\begin{document}

\title{Few-layer Phosphorene: An Ideal 2D Material For Tunnel Transistors}

\author[1*]{\underline{Tarek A. Ameen}}
\author[1*]{\underline{Hesameddin Ilatikhameneh}}
\author[1]{Gerhard Klimeck}
\author[1]{Rajib Rahman}
\affil[1]{\normalsize{Network for Computational Nanotechnology, Department of Electrical and Computer Engineering, Purdue University, West Lafayette, IN 47907, USA}}
\affil[*]{\normalsize{These authors contributed equally to this work.}}
\renewcommand\Authands{ and }
\maketitle
\providecommand{\keywords}[1]{\textbf{\textit{Keywords---}} #1}


\begin{abstract}
2D transition metal dichalcogenides (TMDs) have attracted a lot of attention recently for energy-efficient tunneling-field-effect transistor (TFET) applications due to their excellent gate control resulting from their atomically thin dimensions. However, most TMDs have bandgaps ($E_g$) and effective masses ($m^*$) outside the optimum range needed for high performance. It is shown here that the newly discovered 2D material, few-layer phosphorene, has several properties ideally suited for TFET applications: 1) direct $E_g$ in the optimum range $\sim$1.0-0.4 eV, 2) light transport $m^*$ (0.15$m_0$), 3) anisotropic $m^*$ which increases the density of states near the band edges, and 4) a high mobility. These properties combine to provide phosphorene TFET outstanding $I_{ON}$ 1 mA/um, ON/OFF ratio$\sim$10$^6$, scalability to 6 nm channel length and 0.2 V supply voltage, thereby significantly outperforming the best TMD-TFETs in energy-delay products. Full-band atomistic quantum transport simulations establish phosphorene TFETs as serious candidates for energy-efficient and scalable replacements of MOSFETs. 

\end{abstract}



Metal-Oxide-Semiconductor Field-Effect-Transistors (MOSFETs) have been the workhorse of most modern-day electronics. Although aggressive size scaling of MOSFETs have ushered in an era of ultra-fast miniature electronics, the advantages of scaling are fast disappearing as MOSFETs enter the sub-20 nm regime. In state-of-the-art MOSFETs, direct source to drain tunneling through the channel potential barrier degrades the OFF-state current and causes excessive power dissipation \cite{ionescu}. Tunnel FETs (TFETs) have been proposed to be energy-efficient alternatives to the MOSFET that can reduce the supply voltage ($V_{DD}$) and satisfy the low power requirements in integrated circuits \cite{appenzeller1, appenzeller2}. Although TFETs, in principle, provide a steep OFF to ON transition needed to minimize power dissipation, the ON-currents of TFETs are quite low \cite{sarkar2015subthermionic, choi2007tunneling}, which deteriorates their operational speed and energy-delay product \cite{nikonov2015benchmarking}. The current level in TFETs is the result of band to band tunneling (BTBT) of carriers, and hence, highly sensitive to the effective masses ($m^*$) and bandgaps ($E_g$) of the channel material. While a small $m^*$ and $E_g$ improve the ON-current ($I_{ON}$) and supply voltage scaling, the same also deteriorate the OFF-current and channel length ($L_{ch}$) scaling through direct source-to-drain tunneling \cite{ilatikhameneh2015can}. To meet the simultaneous requirement of the semiconductor industry of both power supply and size scaling, materials need to be carefully chosen with optimized $m^*$ and $E_g$. In this work, it is shown that the newly studied few-layer phosphorene \cite{liu2014phosphorene} provides the ideal material properties to obtain high performance in TFETs as well as to simultaneously achieve both $V_{DD}$ and $L_{ch}$ scaling.

There are several solutions to the low $I_{ON}$ challenge of TFETs \cite{sarkar2015subthermionic}. $I_{ON}$ depends exponentially on $E_g$, $m^*$, and the electric field $F$ at tunnel junction (i.e. $log(I_{ON}) \propto \frac{-\sqrt{m^*E_g}}{F}$).  Hence, $I_{ON}$ can be enhanced either by a) increasing $F$ or by b) using a channel material with optimum $E_g$ and $m^*$. A number of approaches for increasing the electric field $F$ were proposed before such as 1) atomically thin 2D channel materials that provide a tight gate control and small tunneling distance \cite{fiori2014electronics,zhang2014two,sarkar2015subthermionic}, 2) dielectric engineering with high- and low-k spacers \cite{ilatikhameneh2015dielectric, chen2015configurable}, 3) internal polarization in Nitrides \cite{Wenjun1}.

In addition to having an atomically thin channel that improves F, few-layer phosphorene also has the optimum $E_g$ and $m^*$ required for high performance TFETs. Moreover, the bandgap of phosphorene remains direct as the number of layers increases. In this regard, phosphorene has a great advantage over other 2D materials, such as graphene and transition metal dichalcogenides (TMDs). Graphene lacks a bandgap and even with engineered bandgaps, it remains unsuitable for transistor applications \cite{zhang2009direct}. Most monolayer TMDs have a bandgap larger than 1 eV. While the Eg of some multi-layer TMDs may reach below 1 eV, multi-layer TMDs are usually indirect gap materials in which the requirement of momentum change of the carriers by phonons causes very low ON-currents. Among TMDs, {only} WTe$_2$ in 2H phase has a moderate $E_g$ of 0.75 eV, however it suffers from a large $m^*$ \cite{ilatikhameneh2015tunnel}, and the 2H phase of WTe$_2$ has not been experimentally demonstrated yet. Density Functional Theory (DFT) calculations predict that $E_g$ of phosphorene varies from about 1.4 eV in monolayer to 0.3 eV in bulk\cite{qiao2014high}. Also, phosphorene has lighter $m^*$ for both electrons and holes of $\sim$ 0.15 $m_0$. Hence, phosphorene is expected to provide the highest performance among all the 2D material TFETs considered so far. 

\begin{figure}[H]
\centering
        \subfigure[]{\label{fig:a}\includegraphics[width=53mm]{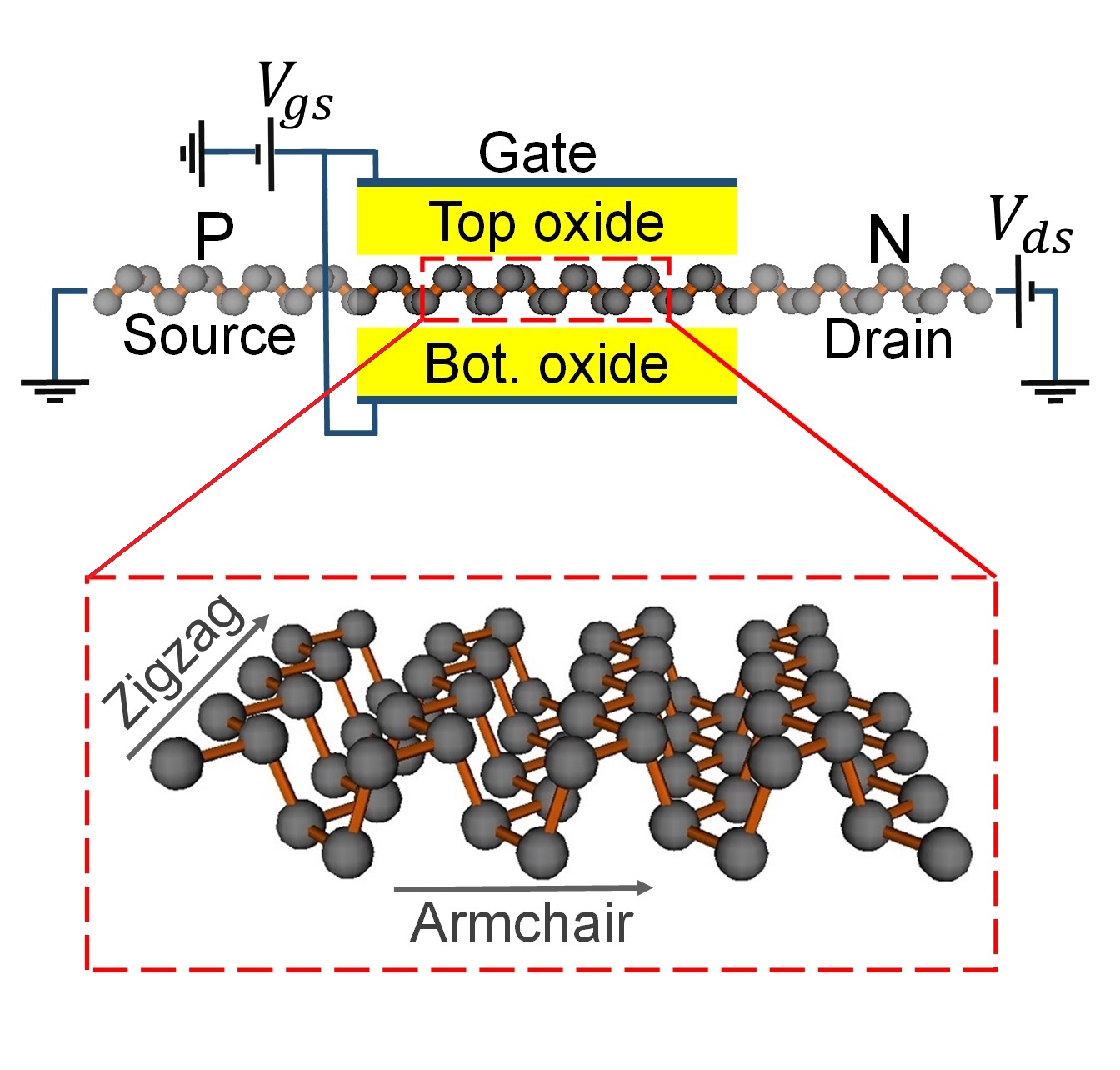}}   
        \subfigure[]{\label{fig:b}\includegraphics[width=53mm]{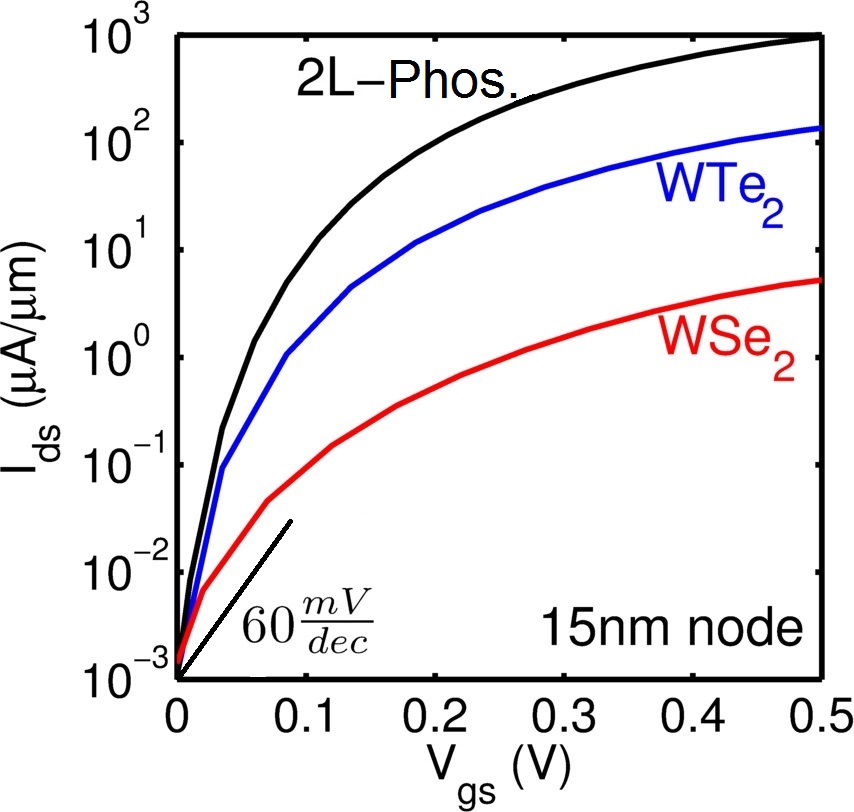}}
        \subfigure[]{\label{fig:c}\includegraphics[width=53mm]{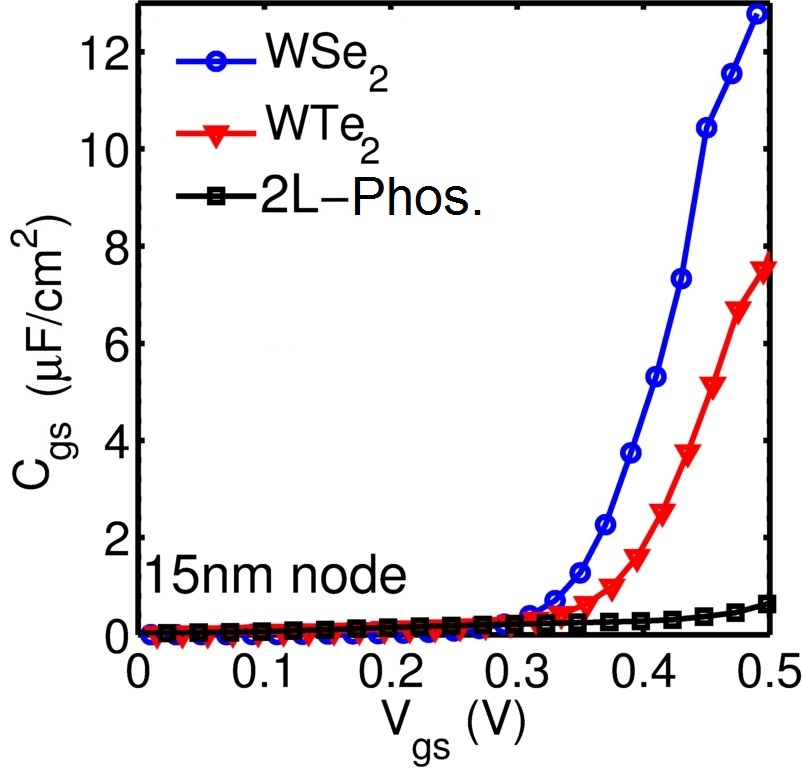}}                     
\caption{\textbf{(a)}: The device structure of a monolayer phosphorene TFET. The channel is oriented along the armchair direction. 
{ \textbf{(b)}: The transfer characteristics ($I_{ds}-V_{gs}$) and \textbf{(c)}: the capacitance voltage ($C_{gs}-V_{gs}$) characteristics of bilayer-phosphornene (2L-phosphorene), monolayer WSe$_2$ and monolayer WTe$_2$ TFETs for $L_{ch}$ of 15 nm and $V_{ds}$ of 0.5 V.}  Phosphorene TFET has 7.5 times higher $I_{ON}$, 4.9 times lower capacitance and 176 times lower intrinsic energy-delay product than WTe$_2$. }
\label{comp_fig}
\end{figure}

In a few-layer phosphorene flake, each layer is a hexagonal honey comb lattice with puckered surface, as shown in Fig. \ref{comp_fig}a. The electron and hole effective masses $m^*$ are highly anisotropic; $m^*$ is low in the armchair direction ($\approx$ 0.15 $m_0$) and is very high in the zigzag direction ($>$ 1 $m_0$)\cite{qiao2014high}. Since the tunneling probability decreases exponentially with the transport effective mass \cite{ilatikhameneh2015fowler}, it is best to have the channel oriented along the armchair direction for high $I_{ON}$. In such a case, the very large $m^*$ in the transverse zigzag direction results in a high density of states near the band edges. This $m^*$ anisotropy ultimately leads to a large $I_{ON}$, as shown later in the paper. The scaling of TFETs to the sub-10 nm regime also require engineering $E_g$ and $m^*$ to keep the ON and OFF state performance intact \cite{ilatikhameneh2015can}. However, to achieve this in most conventional materials such as III-Vs, complicated experimental techniques need to adopted such as application of strain or forming alloys, which can also introduce disorder in the device. In this regard, the layer dependent $E_g$ and $m^*$ in phosphorene already provides an additional knob to optimize the performance for sub-10 nm TFETs, as shown later. 


Experimentally, phosphorene flakes as thin as a single layer have been realized recently by means of mechanical exfoliation \cite{liu2014phosphorene}. The experimental $E_g$ of a single layer phosphorene has been measured to be approximately 1.45 eV which is way higher than the bulk $E_g$ of black phosphorus ($\approx$ 0.3 eV). Measured few-layer phosphorene carrier mobility is very high in the armchair direction, it is $\approx$ 256 cm$^2$/Vs for few-layers and $\approx$ 1000 cm$^2$/Vs for bulk \cite{liu2014phosphorene}. In addition, strong anisotropy of $m^*$ was verified by angle dependent conductivity \cite{liu2014phosphorene}. Later, Saptarshi et al. reported experimental measurements of the thickness dependent transport gap and Schottky barriers of phosphorene \cite{Exp1}. However, there are challenges to the development of phosphorene based electronics as well. Few-layer phosphorene is unstable in atmosphere and is prone to humidity and oxygen. Hence, it degrades within several hours when left in air\cite{castellanos2014isolation,kou2015phosphorene}. However, there are many efforts to solve this stability challenge; e.g. Junhong et al. stabilized phosphorene for two months by encapsulating it within $Al_2O_3$\cite{na2014few}.

In this work, we performed full band atomistic quantum transport simulations of phosphorene TFETs based on the non-equilibrium Green's function simulator NEMO5 with a second nearest neighbor sp$^3$d$^5$s$^*$ tight-binding (TB) Hamiltonian. The electrostatics of the device is obtained by solving a 3D finite-element Poisson equation self-consistently with the quantum transport equations described in the Methods section. The simulated phosphorene TFET assumes a double gated structure as shown in Fig. \ref{comp_fig}a. The channel length is 15 nm and the transport direction is oriented along the armchair direction. The source and drain doping levels are set to $10^{20}$ cm$^{-3}$ in a p-i-n configuration, effective oxide thickness (EOT) is 0.5 nm, and the drain bias $V_{ds}$ equals 0.5 V unless mentioned otherwise. The device specifications are compatible with the international technology road-map for semiconductors (ITRS) projections for 2027 \cite{ITRS}.


Figs. \ref{comp_fig}b and \ref{comp_fig}c compare the current-voltage ($I_{ds}-V_{gs}$) and capacitance-voltage ($C_{gs}-V_{gs}$) characteristics respectively of bilayer-phosphorene (2L-phosphorene) with those of WTe$_2$ and WSe$_2$ (which have been identified as the best TMD material candidates for TFETs \cite{ilatikhameneh2015tunnel}) for a supply voltage $V_{DD}$ of 0.5 V. 2L-phosphorene provides an inverse sub-threshold slope (SS) much lower than the other two TMDs (well below the Boltzmann limit of 60 mV/dec at room temperature), and provides an $I_{ON}$ of nearly 1 mA/um (about 7.5 times higher than WTe$_2$ in 2H phase). The ON-state capacitance of 2L-phosphorene is also about 5 times lower than that of WTe$_2$. The large $I_{ON}$ and small $V_{DD}$ and $C_{gs}$ translate into a very small switching energy and switching delay for the 2L-phosphorene. The most important metric of performance for low power transistors is the product of the switching energy and the delay (energy-delay product or EDP) \cite{nikonov2015benchmarking}. The lower the EDP, the more energy-efficient and faster the device is. 2L-phosphorene has 176 times lower EDP compared to the best TMD TFET (WTe$_2$). The origins of these improvements are discussed next.

\begin{figure}[H]
\centering
        \subfigure[]{\label{fig:a}\includegraphics[width=51mm]{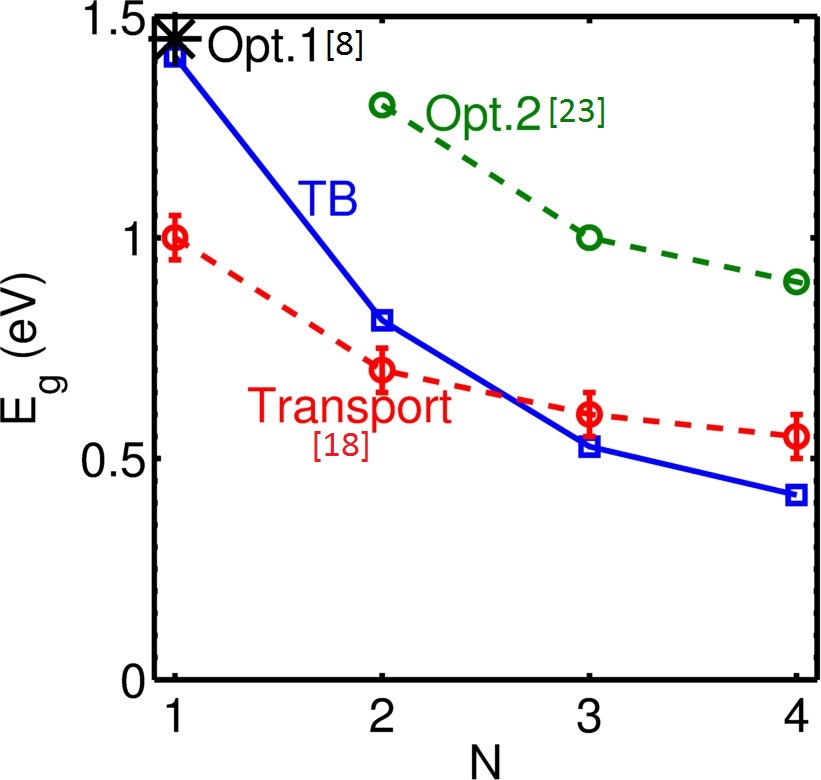}}
        \subfigure[]{\label{fig:b}\includegraphics[width=53mm]{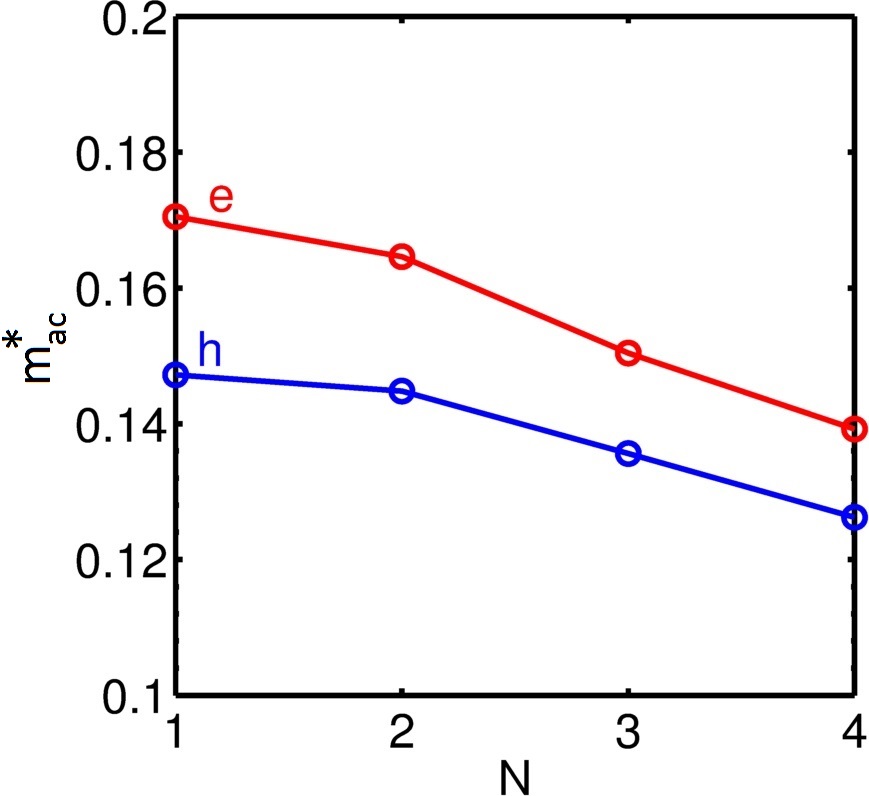}}
        \subfigure[]{\label{fig:b}\includegraphics[width=53mm]{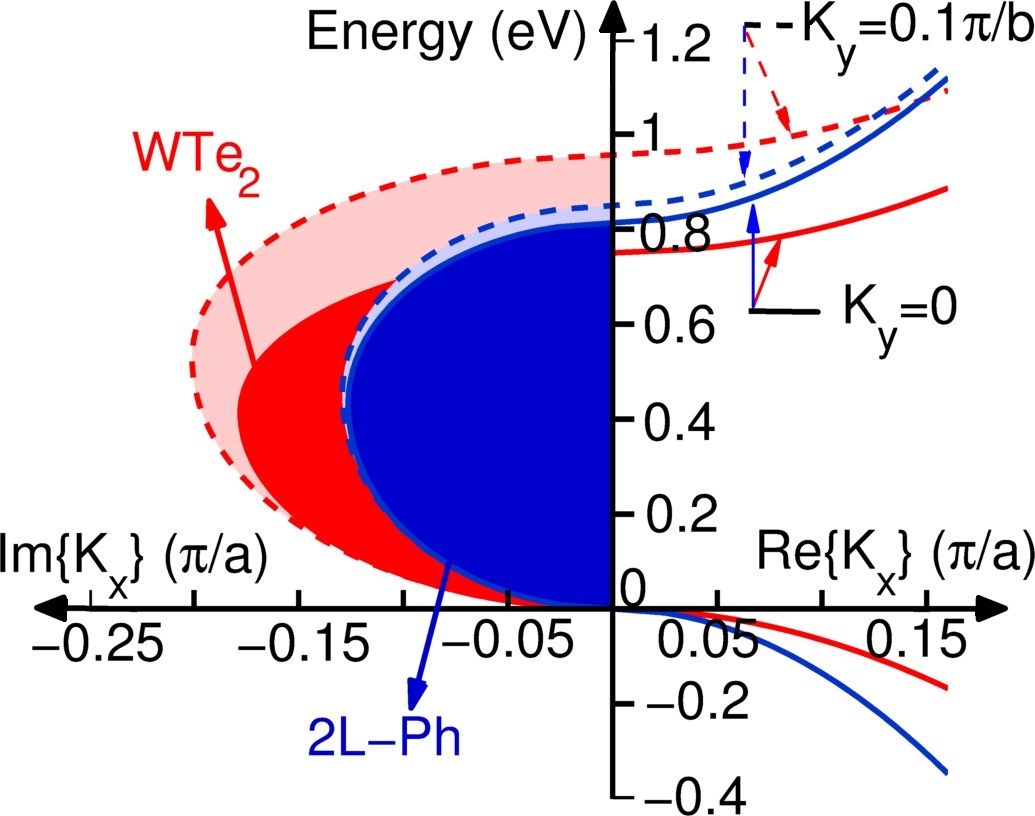}}            
\caption{a) The bandgap $E_g$ and  b) effective masses along armchair direction $m^*_{ac}$ as function of the number of layers N. Bandgaps measured in transport experiments \cite{Exp1} differ from those of optical measurements \cite{liu2014phosphorene, Exp2}, and both are shown as reference. The DFT guided TB bandgaps follow the transport measurements more closely for multi-layer phosphorene. c) The complex band structure of 2L-phosphorene and monolayer WTe$_2$. The complex bands are plotted at transverse wave-vector $K_y = 0$ and $0.1 \pi/b$ for both materials. The area enclosed by the imaginary wave-vector and the vertical axis (i.e. the shaded area) determines the BTBT decay rate.  Bilayer phosphorene not only has smaller BTBT decay rate at $K_y = 0$ due to small transport mass, but also at non-zero $K_y$ due to large transverse $m^*$.}
\label{EHB_fig}
\end{figure}

Figs. \ref{EHB_fig}a and \ref{EHB_fig}b show $E_g$ and $m^*$ in the armchair direction ($m^*_{ac}$) as a function of the number of phosphorene layers extracted from phosphorene bandstructures computed with the atomistic tight-binding model of this work. In Ref. \cite{ilatikhameneh2015can}, optimum $E_g$ and $m^*$ values needed to maximize $I_{ON}/I_{OFF}$ in TFETs were presented for various supply voltages and channel lengths $L_{ch}$. It was suggested that for $L_{ch} = $ 15 nm and $V_{DD} = $ 0.5 V, $E_g$ and $m^*$ need to be roughly about 0.7 V and 0.15 $m_0$ respectively. It is seen in Fig. \ref{EHB_fig}b that the electron and hole $m^*$ are roughly about 0.15 $m_0$ and do not vary much with the number of layers. While the $E_g$ in Fig. \ref{EHB_fig}a is seen to be strongly dependent on the number of layers, apart from the 1.4 eV value for monolayer phosphorene, $E_g$ is mostly in the range of 0.7 to 0.4 eV, with the optimum value of 0.7 eV reached for 2L-phosphorene. It is to be noted that there is still some experimental discrepancy about the actual values of $E_g$ in phosphorene with transport measurements yielding smaller bandgaps than optical measurements (as also seen in TMDs). On the other hand, DFT and DFT-guided TB calculations used in this work yield bandgaps closer to the transport measurements. In spite of these differences, the layer dependent variation of $E_g$ from the bulk value of 0.4 eV suggests enough scope to optimize $E_g$ by varying the channel thickness.    


The bandgap alone does not explain why the phosphorene TFET significantly outperforms WTe$_2$ TFET since 2L-phosphorene has a similar $E_g$ as 1L-WTe$_2$. The difference actually originates from 2L-phosphorene having a light transport $m^*$ in the armchair direction ($m_{ac}^*$) and a heavy transverse $m^*$ in the zigzag direction ($m_{zz}^*$). This is conveniently illustrated in the complex bandstructure in Fig. \ref{EHB_fig}c, which shows the energy-momentum dispersion of the carriers in the forbidden bandgap connecting the conduction and valence band states. The complex part of the bandstructure corresponds to the evanescent wavefunctions $e^{-\kappa z}$ in the bandgap with imaginary momentum $i \kappa$, and the area enclosed by the imaginary band and the energy axis corresponds to the band to band tunneling (BTBT) decay rate \cite{guan2011complex}. The smaller the area, the larger is the transmission probability. Fig. \ref{EHB_fig}c compares the complex band structure of 2L-phosphorene with 1L-WTe$_2$. The complex bands are plotted at transverse wave-vector $K_y = 0$ and $0.1 \pi/b$ for both materials. 2L-phosphorene not only has a smaller BTBT decay rate at $K_y = 0$ (due to small transport $m^*$), but also at a non-zero $K_y$. This is due to a large transverse $m^*$ ($m_{zz}^*$) which prevents the decay rate from increasing significantly with $K_y$. In other words, phosphorene has a high density of states of carriers with optimum transport $m^*$ and $E_g(K_y)$.

\begin{figure}[H]
\centering
        \subfigure[]{\label{fig:a}\includegraphics[width=53mm]{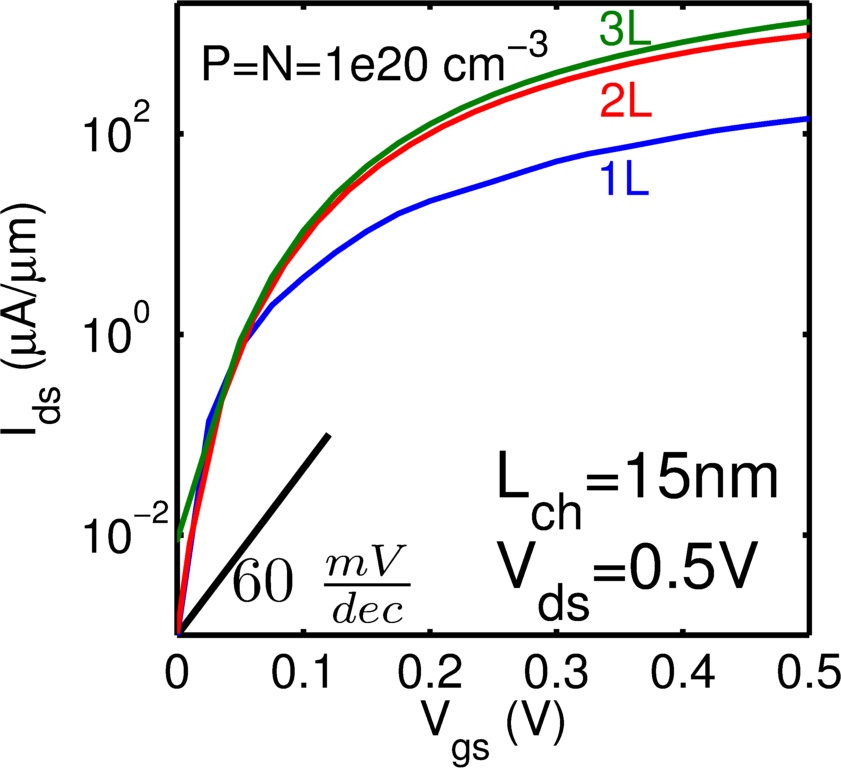}}
        \subfigure[]{\label{fig:b}\includegraphics[width=53mm]{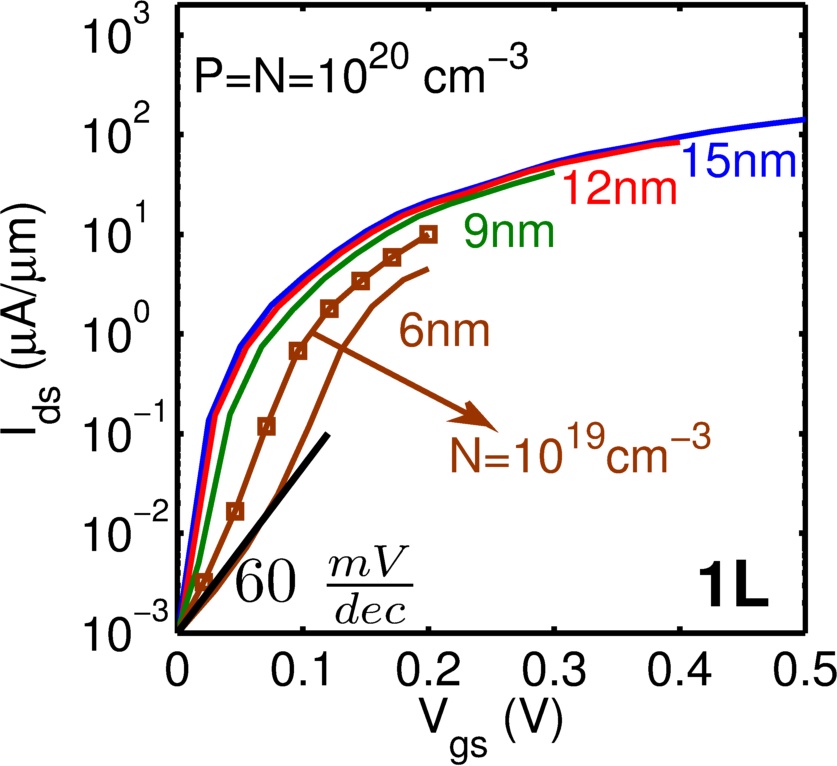}}
        \quad
        \subfigure[]{\label{fig:c}\includegraphics[width=53mm]{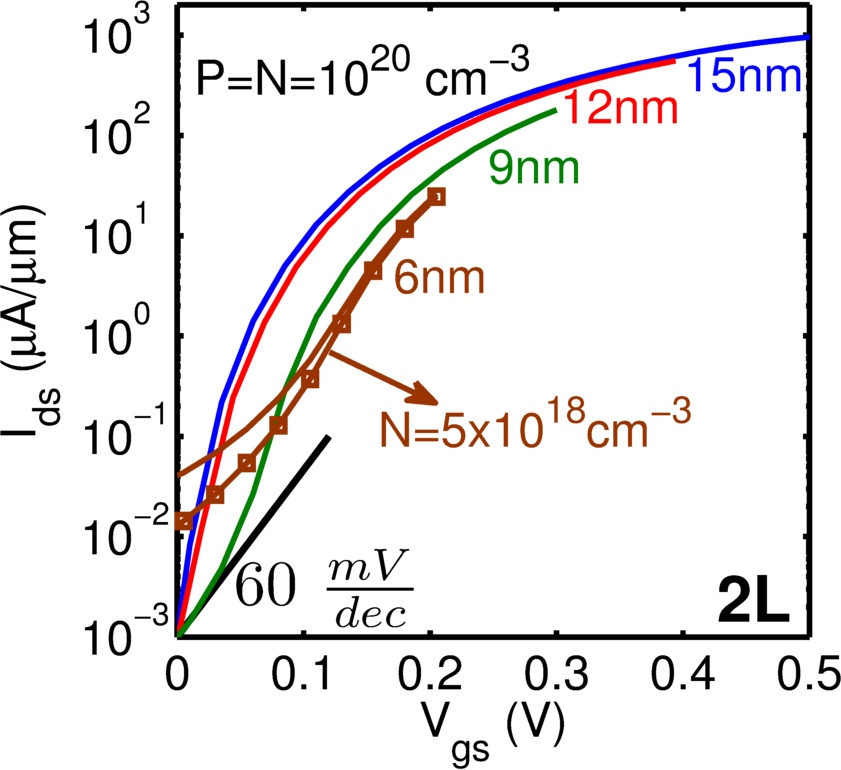}}
        \subfigure[]{\label{fig:d}\includegraphics[width=53mm]{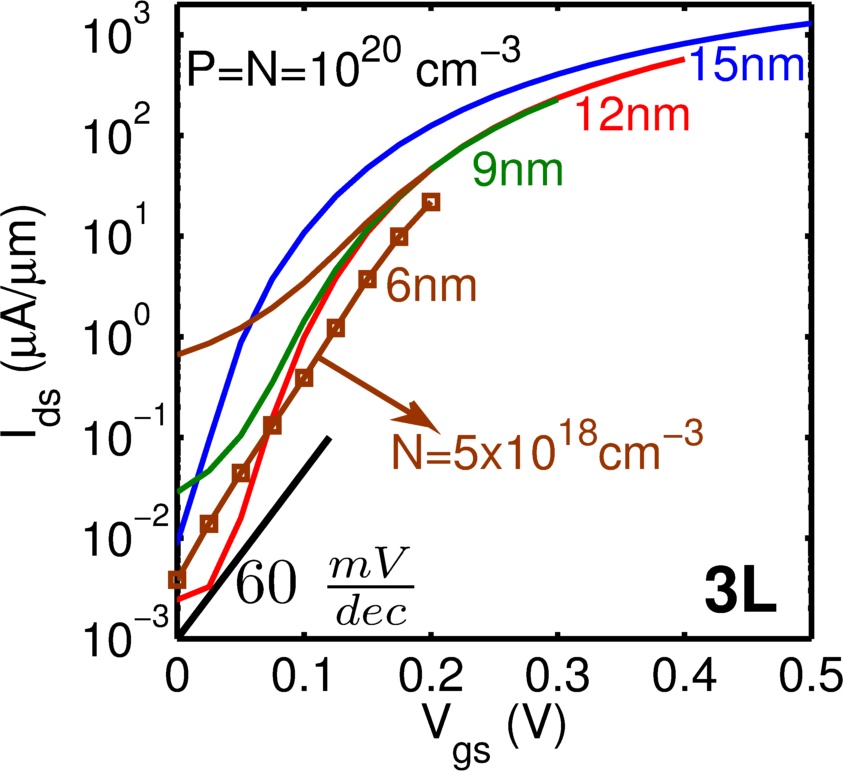}}              
\caption{a) The transfer characteristics of the mono- (1L), bi- (2L), and tri-layer(3L) phosphorene TFETs for 15 nm  channel length $L_{ch}$. Transfer characteristics of constant electric field $E$ scaling (i.e. $E = \frac{V_{DD}}{L_{ch}} = $ 30 V/nm) for (b) 1L, (c) 2L, and (d) 3L phosphorene. For the $L_{ch} = $6 nm case, the $I_{ds}-V_{gs}$ can be optimized through asymmetric doping.}
\label{Scaling_I_fig}
\end{figure}

Next, the performance of the phosphorene TFET and its scalability in $V_{DD}$ and $L_{ch}$ are evaluated as a function of the number of layers. Fig. \ref{Scaling_I_fig}a shows the transfer characteristics of mono- (1L), bi- (2L), and tri-layer (3L) phosphorene TFETs with $L_{ch}$ of 15 nm. The 2L-phosphorene provides the highest ON/OFF current ratio. Notice that although 3L phosphorene provides higher $I_{ON}$, it has a higher $I_{OFF}$ compared to the 2L case. Fig. \ref{Scaling_I_fig}b to \ref{Scaling_I_fig}d show the transfer characteristics of scaled few-layer phosphorene at different technology nodes. Constant electric field $E$ scaling (i.e. $E = \frac{V_{DD}}{L_{ch}}$) of 30 V/nm is considered here. Doping level of source and drain is assumed to be symmetric unless mentioned otherwise. In almost all of the three cases, the phosphorene TFET seems to scale very well from 15 nm to 9 nm channel lengths. Although for very short $L_{ch}$ such as 6 nm, $I_{OFF}$ degrades significantly, asymmetric doping can be used to suppress the p-branch of the TFET and reduce $I_{OFF}$. For the $L_{ch}$ = 6 nm case, reducing the drain doping ($N_d$) increases the drain to channel tunneling distance \cite{Analytic1} and helps to block $I_{OFF}$. However, there is a lower limit to $N_d$. Reducing $N_d$ reduces the carrier density (through $E_c-E_F$) and the tunneling window. For the $L_{ch}$ = 6 nm case, the optimum $N_d$ is found to be $10^{19}$ cm$^{-3}$ in 1L, and $5\times10^{18}$ cm$^{-3}$ in 2L and 3L as shown in Fig. \ref{Scaling_I_fig} b-d). 1L case shows the highest ON/OFF current ratio in the 6 nm case. 

\begin{figure}[H]
\centering
        \subfigure[]{\label{fig:a}\includegraphics[width=53mm]{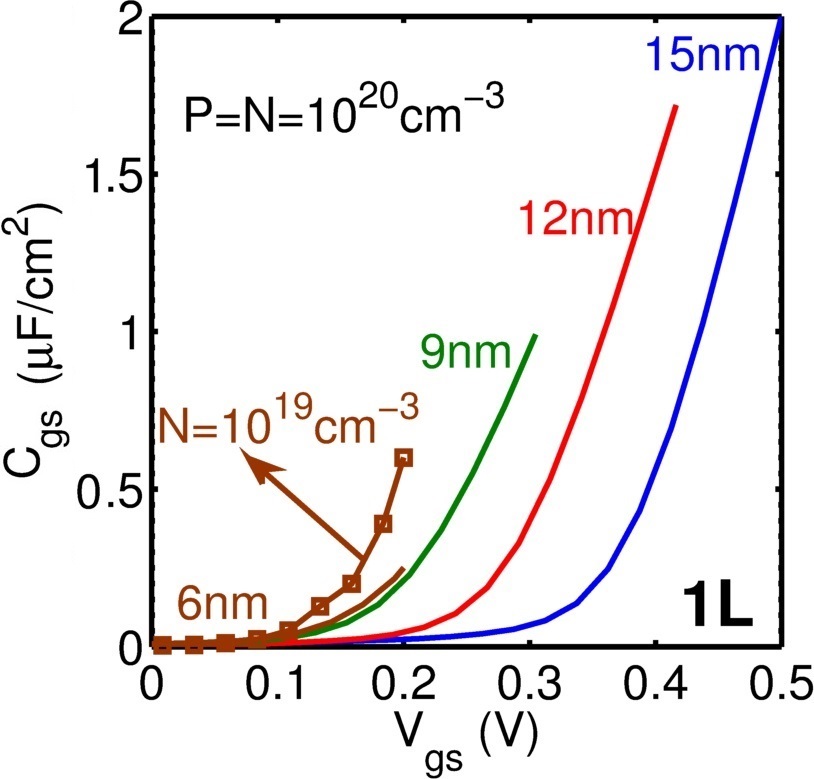}}
        \subfigure[]{\label{fig:b}\includegraphics[width=53mm]{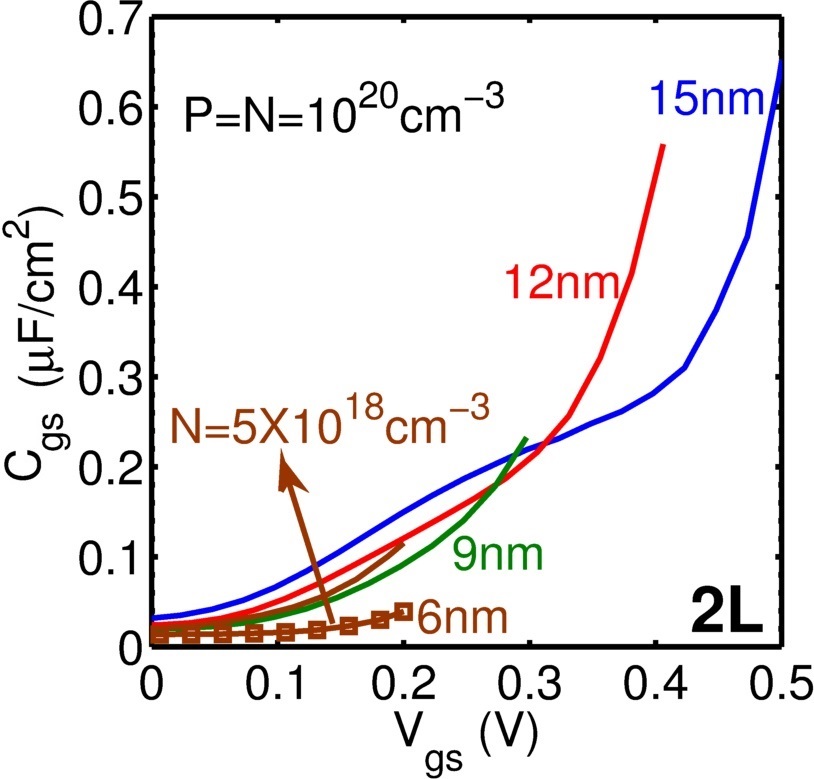}}
        \subfigure[]{\label{fig:c}\includegraphics[width=53mm]{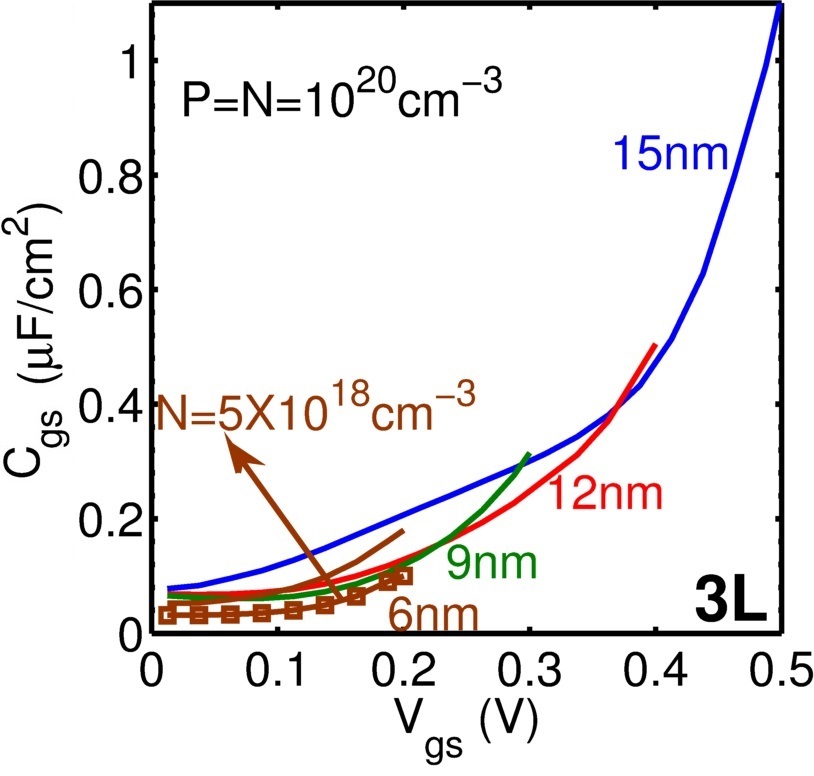}}         
\caption{Capacitance-voltage characteristics of (a) 1L-, (b) 2L- and (c) 3L-phosphorene. Scaling down $V_{DD}$ and $L_{ch}$ reduces the capacitance and improves the transient response. The capacitances of 2L-phosphorene and 3L-phosphorene are lower than half the 1L-phosphorene.}
\label{Scaling_c_fig}
\end{figure}

\begin{figure}[H]
\centering
\includegraphics[width=60mm]{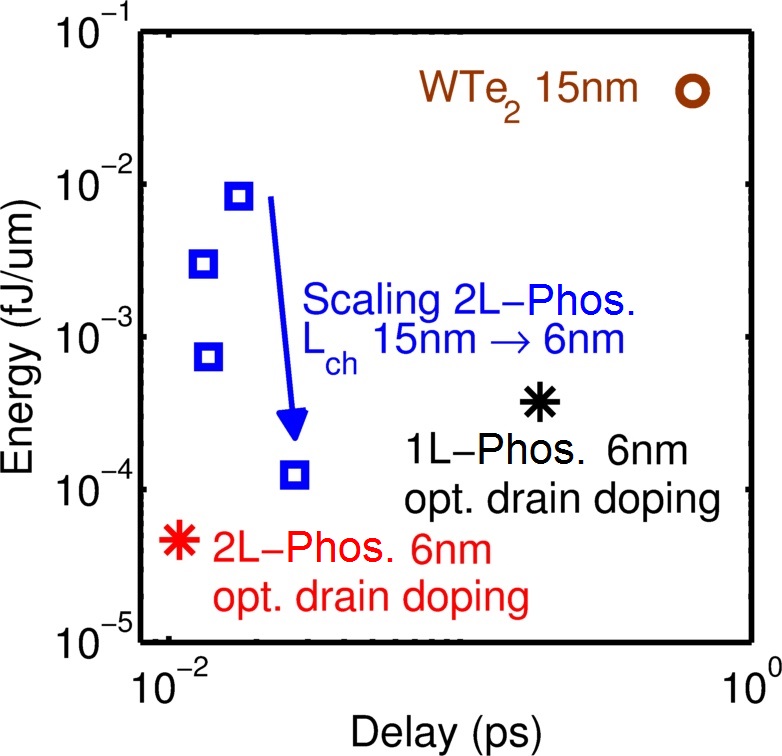}
\caption{The energy delay (ED) of 2L-phosphorene at different scaling nodes. Phosphorene provides lower ED than WTe$_2$. For the 6 nm node, optimized 2L-phosphorene with asymmetric doping provides the lowest possible EDP. }
\label{EDP_fig}
\end{figure}

The total gate capacitances ($C_{gs}$) of 1L- to 3L-phosphorene TFETs are shown in Fig. \ref{Scaling_c_fig} for the same constant electric field scaling discussed before. As expected, the capacitances also scale quite well up to $L_{ch} =$ 9 nm. 2L-phosphorene offers the lowest capacitances. Although the capacitances for the $L_{ch}$ = 6 nm case are slightly larger than the 9 nm case, asymmetric doping can decrease the capacitance for 2L and 3L, as shown Fig. \ref{Scaling_c_fig}. The capacitances predicted here for phosphorene are much less ($< 10\%$) than those reported for TMDs \cite{ilatikhameneh2015tunnel}.
The lower $C_{gs}$ in phosphorene originates from its optimum $E_g$ and $m^*$. The $I_{ds}$-$V_{gs}$ and $C_{gs}$-$V_{gs}$ are shifted in voltage axis such that the current at zero gate voltage $I_{OFF}$ is set to 1 nA/um as required by ITRS \cite{ITRS}. Lower currents in TMDs, which is a result of their higher $E_g$ and $m^*$, makes 0 gate voltage to be closer to threshold voltage if compared with phosphorene. Accordingly, TMDs operate closer to ON-state which results in a higher amount of charge in channel and a higher $C_{gs}$. In summary, the benefits of optimum $E_g$, small transport $m^*$, and large transverse $m^*$ in phosphorene are two-fold: 1) higher $I_{ON}$, and 2) lower capacitance.
The outstanding $I_{ds}$-$V_{gs}$ and $C_{gs}$-$V_{gs}$ characteristics of few-layer phosphorene translate into impressive energy-delay products, which is used ultimately to compare ultra-fast energy-efficient transistors. Fig. \ref{EDP_fig} shows the computed intrinsic energy and delay of phosphorene TFETs compared to WTe$_2$, which has been benchmarked as the best TMD TFET \cite{ilatikhameneh2015tunnel}. It has been shown that WTe$_2$ also outperforms complementary (C) MOS low power FETs in a simulation of the extrinsic delay of a 32 bit full adder \cite{ilatikhameneh2015tunnel}. In a plot of energy versus delay such as Fig. \ref{EDP_fig}, the bottom left corner with the lowest EDP is preferred. It is seen in Fig. \ref{EDP_fig} that the EDPs of phosphorene TFETs are much smaller than the best TMD TFET. EDP of 2L-phosphorene with $L_{ch}$ of 15 nm is two orders of magnitude smaller than the EDP of the WTe$_2$ TFET. Not only does phosphorene provide record $I_{ON}$ and $C_{gs}$ but also a record energy delay product among 2D materials. The optimized asymmetric doping also improves the energy delay product of TFETs specially for sub-9 nm channel lengths.

In conclusion, few-layer phosphorene has a unique set of properties which makes it an excellent candidate for future ultra-scaled low power electronics: 1) atomistically thin body thickness, 2) tune-able $E_g$ and $m^*$ with number of layers within the optimum range for TFET applications, 3) anisotropic $m^*$, and 4) direct band gap even in multi-layer. 
These features make phosphorene an exceptional candidate among 2D materials for TFET applications. The $I_{ds}$-$V_{gs}$ and $C_{gs}$-$V_{gs}$ characteristics of few-layer phosphorene exhibit orders of magnitude improvement in energy-delay product compared to other 2D TFETs (e.g. TMD TFETs). Bilayer phosphorene shows optimum performance and is recommended for adoption as the future material of 2D-TFETs. 

\section*{Methods}
In the quantum transport simulations performed in this work, the phosphorene Hamiltonian employs a 10 band sp$^3$d$^5$s$^*$ 2nd nearest neighbor tight binding model (TB). The TB parameters have been optimized to reproduce the band structures obtained from density functional theory (DFT) using HSE06. A general TB parameter set was obtained that captured the bandstructure of monolayer to bulk phosphorene. This DFT to TB mapping is a standard technique in semi-empirical TB \cite{tan2015tight}. The Hamiltonian is represented with TB instead of DFT, since DFT is computationally expensive and is size limited. Our TB model agrees well with previous calculations of $m^*$ and $E_g$ of few-layer phosphorene obtained from DFT with Becke Johnson functional (DFT-mBJ)\cite{qiao2014high} 

For transport simulations, a self- consistent Poisson-quantum transmitting boundary method (QTBM) has been used with the TB Hamiltonian. QTBM is equivalent to the non equilibrium Green's function (NEGF) approach in the ballistic case, but it entails the solution of a linear system of equations instead of obtaining the Green's function by matrix inversion which is more computationally inefficient \cite{luisier2006atomistic}. In QTBM, the Schroedinger equation with open boundary conditions is given as, 
\begin{equation}
(EI-H-\Sigma) \Psi_{S/D} = S_{S/D},
\label{eqn_1}
\end{equation}
 where $E$, $I$, $H$, and $\Sigma$ are the carrier energy, identity matrix, device Hamiltonian, and self-energy due to open boundaries and $\Psi$ and S are the wave function and a carrier injection term respectively from either source (S) or drain (D). 3D Poisson equation is solved using the finite-element method. It should be noted that the dielectric tensor $\varepsilon$ of few-layer phosphorene is anisotropic and has been obtained from DFT calculations \cite{wangprediction}. 
The Poisson equation reads as follows :
\begin{equation}
\frac{d}{dx}\left(\varepsilon_x\frac{dV}{dx}\right)+\frac{d}{dy}\left(\varepsilon_y\frac{dV}{dy}\right)+\frac{d}{dz}\left(\varepsilon_z\frac{dV}{dz}\right) = - \rho,
\label{eqn_2}
\end{equation}
where $V$ and $\rho$ are the electrostatic potential and total charge, respectively. In this paper, the transport simulations have been performed with the Nanoelectronics Modeling tool NEMO5\cite{nemo5,nemo5_2}.

\bibliographystyle{ieeetr}
\bibliography{thesis}

\section*{Acknowledgment}
This work was supported in part by the Center for Low Energy Systems Technology (LEAST), one of six centers of STARnet, a Semiconductor Research Corporation program sponsored by MARCO and DARPA. nanoHUB.org computational resources are used. 
operated by the Network for Computational Nanotechnology funded by the US National Science Foundation under grant EEC-1227110, EEC-0228390, EEC-0634750, OCI-0438246, and OCI-0721680 is gratefully acknowledged.
\section*{Author Contributions}
T.A and H. I. performed the simulations and analyzed the data. G. K. and R.R. supervised the work. All authors contributed to writing the manuscript.

\subsection*{Competing financial interests}
The authors declare no competing financial interests.

\end{document}